\documentclass[prd,aps,nofootinbib,floatfix,11pt]{revtex4}
\usepackage{amsmath,graphicx,epsfig,amssymb,dsfont,mathtools,}
\usepackage[usenames]{color}
\usepackage{ulem} 
\usepackage{bigstrut}
\usepackage{slashed}
\usepackage{multirow}
\usepackage{subfigure}

\allowdisplaybreaks

\begin{document}\title{Hadronization of heavy diquark and light quark within NJL-Model}
\author{Yu-Ji Shi~\footnote{Email:\ shiyuji92@126.com}}
\affiliation{ Helmholtz-Institut f\"ur Strahlen- und Kernphysik and Bethe Center \\ for Theoretical Physics,
  Universit\"at Bonn, 53115 Bonn, Germany}

\begin{abstract}
We present a path-integral hadronization for doubly heavy baryons. The two heavy quarks in the baryon are approximated as a scalar or axial-vector diquark described by a heavy diquark effective theory. The gluon dynamics are represented by a NJL-Model interaction for the heavy diquarks and light quarks, which leads to an effective action of the baryon fields after the quark and diquark fields are integrated out. This effective action for doubly heavy baryon includes the electromagnetic and electroweak interactions, as well as the interaction with light mesons. We also verify the Ward-Takahashi identity at the baryon level, obtain the Isgur-Wise function for weak transitions, and calculate the strong coupling constant of the doubly heavy baryon and pion. Numerical studies are also performed.
\end{abstract}
\maketitle

\section{Introduction}
In the past few decades, the traditional quark model has successfully explained various of
hadronic states observed from the experiments. However, there still remains some exotic particles predicted by the quark model has not been experimentally observed or established. One of such particles is the doubly heavy baryon, which is a baryonic state made up of two heavy and one light quarks. After years of searching, in the light of the great prediction \cite{Yu:2017zst}  the lowest-lying doubly heavy baryon $\Xi_{cc}$ was finally observed by the LHCb collaboration in 2017~\cite{Aaij:2017ueg}, with its mass being $m_{\Xi_{cc}^{++}}=3621.40\ {\rm MeV}$.
This inspiring observation encourages people to believe that more heavier doubly heavy baryons will
be observed through the continuous experimental researches~\cite{Traill:2017zbs,Cerri:2018ypt,Aaij:2019jfq} in the future. On the theoretical side, people are trying to understand the dynamical and spectroscopical
properties of the doubly-heavy baryon states, see e.g. Refs.~\cite{Fleck:1989mb,Wang:2017mqp,Wang:2017azm,
Gutsche:2017hux,Li:2017pxa,Guo:2017vcf,Xiao:2017udy,Sharma:2017txj,Ma:2017nik,Hu:2017dzi,Shi:2017dto,Yao:2018zze,Yao:2018ifh,
Ozdem:2018uue,Ali:2018ifm,Zhao:2018mrg,Wang:2018lhz,
Liu:2018euh,Xing:2018lre,Dhir:2018twm,Berezhnoy:2018bde,Jiang:2018oak,Zhang:2018llc,Li:2018bkh,Gutsche:2018msz,Shi:2019hbf,Shi:2019fph,Hu:2019bqj,Gutsche:2019iac,Brodsky:2011zs,Yan:2018zdt,Hu:2020mxk,Rahmani:2020pol,Ebert:2004ck,Gershtein:2000nx,Kiselev:2017eic,Olamaei:2020bvw,Ozdem:2019zis,Wang:2019dls,Cheng:2020wmk,Gerasimov:2019jwp,Yu:2019lxw,Grossman:2018ptn}. However, a comprehensive description of these properties is still far from complete.

Although understanding the structure of doubly heavy baryon is a great challenge, the situation can be simplified if one reduce the doubly heavy baryon into a two-body system by treating the two heavy quarks as a point-like spin-0 or spin-1 diquark. The idea of diquark has been widely used in the earlier works \cite{Georgi:1990ak,Carone:1990pv,Flynn:2007qt,Nguyen:1993dw}, and it is indeed a reasonable approximation. As argued by Refs.~\cite{Brodsky:2011zs,Yan:2018zdt,Hu:2005gf,Bodwin:1994jh}, in a doubly heavy baryon, the spatial size of the two heavy quarks is at the order of $r_{QQ}\sim 1/{m_Qv}$, while the distance between one of the heavy quarks and the light quark is at the order of $r_{Qq}\sim 1/{\Lambda_{QCD}}$, further, if the heavy quark is heavy enough, one has $m_Qv^2\ll m_Qv\ll m_Q$. Therefore, the small ratio $r_{QQ}/r_{Qq}\sim 1/m_{Q}v \ll 1$ validates the diquark approximation in the heavy quark limit, which enables people to construct effective theories for heavy diquarks \cite{Carone:1990pv,Flynn:2007qt,Nguyen:1993dw,Hu:2005gf,An:2018cln,Shi:2020qde,Soto:2020pfa}.

How to transform a theory of light quarks, heavy diquarks and gluon to an effective theory containing doubly heavy baryon is another challenge. 
Generally,  such transformation is extremely difficult to be realized from the first principle. One of the applicable approaches is to construct a bottom-up type effective theory, for instance the chiral perturbation theory, where the hadron level Lagrangian are built according to the underlying symmetries. Another one is the top-down approach with additional assumptions or approximations being required. Path-integral hadronization belongs to the second class \cite{Cahill:1988zi,Reinhardt:1989rw,Ebert:1992zq,Ebert:1995fp,Ebert:1996ab,AbuRaddad:2002pw}, through which one should firstly introduce some auxiliary fields to represent the expected hadron fields, and then systematically integrate out all the fundamental degrees of freedom such as quarks, diquarks and gluons. 

Obviously, due to the nonlinear gluon self-interactions,  this functional integration is almost impossible to be performed analytically. However, in the literatures, there are two major approaches aiming to overcome this difficulty. One of which is to expand the generating functional of the full theory in terms of the quark color currents \cite{Ebert:1994mf}. Another one is the so-called field strength approach where the original gluon sector is reformulated by a new  field strength Ref.~\cite{Reinhardt:1991sh,Schaden:1989pz,Ebert:1992jz}. Both of the two approaches will finally lead to an effective quark theory in the absence of gluon, which is just the well known NJL-Model \cite{Nambu:1961tp,Nambu:1961fr}. In this work, we will perform a path-integral hadronization with a NJL-Model typed interaction for the heavy diquarks and light quarks.

This article is organized as follows: In section~II, we will perform the hadronization with the approach of path-integral, which will produce an effective action of doubly heavy baryon. With suitable field renormalization, we will obtain the residual mass of the doubly heavy baryons. In section~III, we will derive the effective electromagnetic interaction, and prove the Ward-Takahashi identity in the hadron level. We will then calculate the Isgur-Wise function for doubly heavy baryon transition matrix element in the heavy diquark limit, as well as its strong coupling with pion. Section~IV contains all the numerical studies. Section~V gives the conclusions.

\section{Hadronization for doubly heavy baryon}
\label{sec:PIHadroni}
\subsection{Heavy diquark effective theory}

In this section, we will introduce the Path-integral hadronization for doubly heavy baryon. Generally, for a hadron composed of several quarks, the main idea of the hadronization is to introduce some auxiliary fields as hadron fields, and then integrating out all the quark degrees of freedom. This will leave us an effective action totally in terms of the hadron fields. Practically, in the case of doubly heavy baryon, one can treat the two heavy quarks in the baryon as a diquark. As a result, the three-body system is simplified to a two-body system and what left to us is to hadronize    the heavy diquark and light quark fields. 

The heavy diquark effective theory (HDiET) at leading power was constructed in our previous work \cite{Shi:2020qde}, the effective Lagrangian for the original scalar and axial-vector diquark reads as 
\begin{align}
{\cal L}_{DiET}=\frac{1}{2}m_X {\rm Tr}\left[\bar K (i \slashed \partial -m_X) K\right]+\frac{1}{2} m_X {\rm Tr}[\bar K J_{1} K]+\frac{1}{2} m_X {\rm Tr}[\bar K K J_2^{T}],
\end{align}
with $K$ a multiplet of scalar and axial-vector diquarks
\begin{align}
K=\frac{i\slashed \partial+m_X}{2m_X}(X_{\mu}\gamma^{\mu}+S\gamma_{5})C,~~~~S=\begin{pmatrix} 0& S_{bc}\\
-S_{bc} &0 
\end{pmatrix},\  
X_{\mu}=\begin{pmatrix}X_{bb\mu} & X_{bc\mu}\\
X_{bc\mu} & X_{cc\mu}
\end{pmatrix}.
\end{align}
$S$ and $X_{\mu}$ are the scalar and axial-vector diquark fields in the heavy flavor $\rm{SU}(2)$ representation, both of which are assumed to have the same mass $m_X$. The trace acts in both spinor and flavor space, $C$ is the charge conjugating matrix. $J_{1,2}$ are the external sources $J_{i}={ E}^{\mu}\gamma_{\mu}-\lambda_{i}A^{\mu}\gamma_{\mu}\gamma_5$, which contains an axial-vector field $A_{\mu}=A_{\mu}^{a}T^{a}$ for weak interaction, and an electromagnetic field $E_{\mu}={\cal Q}_{h}A_{\mu}^{em}$. $\lambda_1, \lambda_2$ are two coupling constants for the underlying electroweak interactions, $T^a$ is the $\rm{SU}(2)$ generator and ${\cal Q}_{h}={\rm diag}\{-1/3,2/3\}$ is the electric charge matrix of $b,c$ quarks.

In the heavy diquark limit, redefining the diquark field: $X^{\mu}={\rm exp}[-im_X v\cdot x]X_v^{\mu}$, and the same to $S$,  all the derivatives in the above Lagrangian can be replaced by the baryon velocity $v$. Thus the Lagrangian at heavy diquark and light quark level is simplified as
\begin{align}
{\cal L} &={\cal L}_{Kin}+{\cal L}_{EM}+{\cal L}_{EW},\nonumber\\
{\cal L}_{Kin}  &=\bar{q}(i\slashed D_{em}-m)q+im_{X}S_{ij}^{\dagger}v\cdot\partial S_{ji}-im_{X}X_{ij\mu}^{\dagger}v\cdot\partial X_{ji}^{\mu},\nonumber\\
{\cal L}_{EM}  &=-2m_{X}X_{\mu ij}^{\dagger}g^{\mu\nu}(v\cdot \tilde{E})_{ji,lk}X_{\nu kl}+2m_{X}S_{ij}^{\dagger}(v\cdot \tilde{E})_{ji,lk}S_{kl},\nonumber\\
{\cal L}_{EW} &=m_{X}S_{ij}^{\dagger}\tilde{B}_{ji,lk}^{\mu}X_{\mu,kl}+m_{X}X_{ij\mu}^{\dagger}\tilde{B}_{ji,lk}^{\mu}S_{kl}+2i\ m_{X}\epsilon_{\alpha\beta\nu\mu}X_{ij}^{\dagger\mu}v^{\alpha}\tilde{A}_{ji,lk}^{\beta}X_{kl}^{\nu},\label{Lagran01}
\end{align}
where $D_{\mu}^{em}=\partial_{\mu}-iE_{\mu}^{q}$, $E_{\mu}^{q}={\cal Q}_{l}A_{\mu}^{em}$ with ${\cal Q}_{l}={\rm diag}\{2/3,-1/3\}$ being the $u, d$ quark charge matrix. The flavor indexes of $q$ are omitted while $i,j,k\cdots$ denote heavy flavor indexes. ${\cal L}_{Kin}$ is the kinematic Lagrangian for light quark and the heavy diquark at the leading power of ${\cal O}(1/m_X)$. In this work the velocity label for the heavy diquark fields are omitted, and only the leading power contribution is considered.  In Eq.~(\ref{Lagran01}) we have also defined the following operators
\begin{align}
\tilde{E}_{ji,lk}^{\mu} & =\frac{1}{2}\delta_{il}E_{jk}^{\mu}+\frac{1}{2}\delta_{jk}E_{il}^{\mu},\nonumber\\
\tilde{A}_{ji,lk}^{\mu} & =\frac{\lambda_{1}}{2}\delta_{il}A_{jk}^{\mu}+\frac{\lambda_{2}}{2}\delta_{jk} A_{il}^{\mu},\nonumber\\
\tilde{B}_{ji,lk}^{\mu} & =\frac{\lambda_{1}}{2}\delta_{il}(2A^{\mu}-v^{\mu}v\cdot A)_{jk}+\frac{\lambda_{2}}{2}\delta_{jk}(2A^{\mu}-v^{\mu}v\cdot A)_{il},
\end{align}
where $\tilde{E}_{ji,lk}$ contains EM sources while $\tilde{A}_{ji,lk}, \tilde{B}_{ji,lk}$ contain EW sources.

\subsection{Path-integral hadronization within the NJL-Model}
To avoid the nonlinear sector of gluon, as mentioned previously, the gluon sector can be reformulated by a NJL-model typed interaction for color currents, which has the form
\begin{align}
{\cal L}_{NJL} =-\kappa\  j_{\mu}^{A}j^{A\mu},
\end{align}
with $\kappa$ being the coupling constant, and $A$ is color index. The road to NJL model only depends on the gluon dynamics so that its form of current-current combination is blind to whatever material currents coupling with gluon. Thus, although in the original NJL model, $j_{\mu}^{A}$ is the color current of quark, in our case $j_{\mu}^{A}$ can also contain the color current of diquark, which reads
\begin{align}
j_{\mu}^{A} =m_{X}g\ v_{\mu}\ {\rm tr}\left[S^{\dagger}\bar{t}^{A}S-X_{\alpha}^{\dagger}\bar{t}^{A}{\cal T}^{\alpha\beta}X_{\beta}\right]+\bar{q}\gamma_{\mu}t^{A}q,
\end{align}
with ${\cal T}^{\alpha\beta}=g^{\alpha\beta}-v^{\alpha}v^{\beta}$ a transverse projection operator, which guarantees the identity $v\cdot X=0$. $\bar{t}^{A}=-(t^{A})^T$ is the generator of the $\bar 3$ representation of color $\rm{SU}(3)$, and $g=g_d/g_s$ with $g_d$ the strong coupling constant of the diquark and gluon. Using the color Fierz identity, and only retaining the terms with $Sq$ and $X_{\mu}q$ being color singlet, we have
\begin{align}
{\cal L}_{NJL}=G_{1}{\rm tr}\left[\bar{q}S^{\dagger}\slashed vSq\right]-G_{1}{\rm tr}\left[\bar{q}X_{\alpha}^{\dagger}{\cal T}^{\alpha\beta}\slashed v X_{\beta}q\right],\label{Lagran2}
\end{align}
with $G_{1}=(\kappa/2)m_{X}g$. Combining Eq.~(\ref{Lagran01}) and Eq.~(\ref{Lagran2}), one can perform the hadronization procedure according to the generating functional
\begin{align}
Z[0]=\int{\cal D}q{\cal D}\bar{q}{\cal D}S{\cal D}X_{\mu}{\rm exp}\left[i\int d^{4}x\left({\cal L}_{Kin}+{\cal L}_{EM}+{\cal L}_{EW}+{\cal L}_{NJL}\right)\right].\label{genetateFunc}
\end{align}

The quadratic forms of $Sq$ and $X_{\mu}q$ in ${\cal L}_{NJL}$ can be linearized by introducing additional Gaussian type integration for two auxiliary fermion fields, which are a spinor field $T$ and a spinor-vector field $H_{\mu}$ respectively. These two auxiliary fields are expected to be the doubly heavy baryon fields with its heavy sector in spin-0 or spin-1 state, and will finally form an effective action. Due to the operator ${\cal T}_{\alpha\beta}$ appearing in the Eq.~(\ref{Lagran2}), $H$ must satisfy the the same transverse condition $v\cdot H=0$ as that of $X$. This linearization procedure reads
\begin{align}
 & {\rm exp}\left(i\int d^{4}x\ {\cal L}_{NJL}\right)=\int{\cal D}T{\cal D}H_{\mu}\times \nonumber\\
&{\rm exp}\left[i\int d^{4}x\left(-\frac{1}{G_{1}}\bar{T}_{ij}\slashed vT_{ji}+S_{ij}^{\dagger}\bar{q}T_{ji}+\bar{T}_{ij}qS_{ji}+\frac{1}{G_{1}}\bar{H}_{ij}^{\mu}\slashed vH_{ji\mu}+X_{ij}^{\dagger\mu}\bar{q}H_{ji\mu}+\bar{H}_{ij}^{\mu}qX_{ji\mu}\right)\right].\label{linearizeSqXq}
\end{align}
Note that besides two heavy flavor indexes, both $T$ and $H$ should also have one extra light flavor index which have been omitted for simplicity. For example,  $\bar{T}_{ij}qS_{ji}$ should be indeed written as $\bar{T}_{ij}q^aS_{ji}^a$, with $a$ denoting $u, d$ flavors. Inserting Eq.~(\ref{linearizeSqXq}) into Eq.~(\ref{genetateFunc}), and then performing the integration of light quark fields $q$ and $\bar q$, one arrives at 
\begin{align}
Z[0] =&\int{\cal D}T{\cal D}H_{\mu}{\cal D}S{\cal D}X_{\mu}\ {\rm Det}[i\slashed D_{em}-m]\nonumber\\
&\times{\rm exp}\left\{ i\int d^{4}x\left[\left(S_{ij}^{\dagger},X_{\mu ij}^{\dagger}\right)\begin{pmatrix}{\cal A}_{ji,lk} & {\cal B}_{ji,lk}^{\nu}\\
{\cal C}_{ji,lk}^{\mu} & {\cal D}_{ji,lk}^{\mu\nu}
\end{pmatrix}\begin{pmatrix}S_{kl}\\
X_{\nu,kl}
\end{pmatrix}-\frac{1}{G_{1}}\bar{T}_{ij}\slashed v T_{ji}+\frac{1}{G_{1}}\bar{H}_{ij}^{\mu}\slashed v H_{ji\mu}\right]\right\} \nonumber\\
  =&\int{\cal D}T{\cal D}H_{\mu}\ {\rm Det}[i\slashed D_{em}-m]\ {\rm exp}\left[-{\rm Tr}\ {\rm log}{\cal A}-{\rm Tr}\ {\rm log}\left({\cal D_{\mu\nu}}-{\cal C_{\mu}}{\cal A}^{-1}{\cal B_{\nu}}\right)\right]\nonumber\\
 &\times {\rm exp}\left[i\int d^{4}x\left(-\frac{1}{G_{1}}\bar{T}_{ij}T_{ji}+\frac{1}{G_{1}}\bar{H}_{ij}^{\mu}H_{ji\mu}\right)\right].\label{BaryoneffAction}
\end{align}
The determinant ${\rm Det}[i\slashed D_{em}-m]$ comes from the integration of light quark fields, which only contributes to the self energy correction on photon. Also note that the generating functional in Eq.~(\ref{BaryoneffAction}) is exactly equivalent with the original one in Eq.~(\ref{genetateFunc}). For the baryons fields $T, H_{\mu}$, one can use the following projection operators to divide them into two parts
\begin{equation}
T=\frac{1+\slashed v}{2}T+\frac{1-\slashed v}{2}T=T_{+}+T_{-},
\end{equation}
and similarly for $H_{\mu}$. Note that $T_+$ and $T_{-}$ satisfy $\slashed v T_+=T_+$ and $\slashed v T_-=-T_-$ respectively, which correspond to equations of motion of a heavy baryon and a heavy anti-baryon. Since in this work we do not care about the case of anti-baryons so we only keep the plus terms $T_+, H_+^{\mu}$ in the above action. For simplicity, the subscript can be omitted and the conditions $\slashed v T=T$ and $\slashed v H_{\mu}=H_{\mu}$ are always required.
The four operators in the trace are defined as
\begin{align}
{\cal A}_{ji,lk}  =&\  m_{X}\delta_{jk}\delta_{il}i v\cdot\partial+2m_{X}(v\cdot\tilde{E})_{ji,lk}-T_{ji}^{b}\left(i\slashed D_{em}^{\prime}+m\right)_{ab}^{-1}\bar{T}_{lk}^{a}~,\nonumber\\
{\cal B}_{ji,lk}^{\nu} =& -T_{ji}^{b}\left(i\slashed D_{em}^{\prime}+m\right)_{ab}^{-1}\bar{H}_{lk}^{\nu,a}+m_{X}\tilde{B}_{ji,lk}^{\mu}~,\nonumber\\
{\cal C}_{ji,lk}^{\mu} =& -H_{ji}^{\mu,b}\left(i\slashed D_{em}^{\prime}+m\right)_{ab}^{-1}\bar{T}_{lk}^{a}+m_{X}\tilde{B}_{ji,lk}^{\mu}~,\nonumber\\
{\cal D}_{ji,lk}^{\mu\nu} =& -m_{X}g^{\mu\nu}\delta_{jk}\delta_{il}i v\cdot\partial-2m_{X}g^{\mu\nu}(v\cdot\tilde{E})_{ji,lk}-H_{ji}^{\mu,b}\left(i\slashed D_{em}^{\prime}+m\right)_{ab}^{-1}\bar{H}_{lk}^{\nu,a}\nonumber\\
&+2m_{X}\epsilon_{\beta}^{\ \nu\alpha\mu}iv_{\alpha}\tilde{A}_{ji,lk}^{\beta}~,\label{fourOpers}
\end{align}
where $D_{\mu}^{\prime em}=\partial_{\mu}+iE_{\mu}^{q}$, and $a,b$ denote spinor indexes. After integrating out the diquark fields in the second step of Eq.~(\ref{BaryoneffAction}), we finally arrive at an expected effective action where only baryon fields appear in the functional integration.

\begin{figure}
\includegraphics[width=0.45\columnwidth]{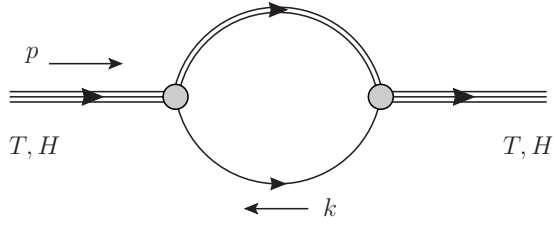} 
\caption{Self energy correction for the baryon fields $T$ or $H$. The triple line denotes the doubly heavy baryon, the double line denotes the heavy diquark, while the gray bubble denotes the interaction between the baryon, heavy diquark and light quark.}
\label{fig:doubBaryonSelfE} 
\end{figure}
The baryon fields should be renormalized according to the coefficients of their quadratic terms in Eq.~(\ref{genetateFunc}). For the $T$ field, they are from the trace term $-(1/i){\rm Tr}\ {\rm log}{\cal A}$ and $-(1/G_1) {\bar T}_{ij}T_{ji}$. The trace term corresponds to the diagram in Fig.~\ref{fig:doubBaryonSelfE} , which leads to
\begin{align}
& -\frac{i}{m_{X}}{\rm Tr}\left[(iv\cdot\partial)^{-1}T_{ji}^{b}(i\slashed\partial+m)_{ab}^{-1}\bar{T}_{lk}^{a}\right]\nonumber\\
  =& \ N_{c}\frac{i}{m_{X}}\int\frac{d^{4}p}{(2\pi)^{4}}{\rm tr}\left[\bar{T}(p)\int\frac{d^{4}k}{(2\pi)^{4}}\frac{\slashed k-m}{(k^{2}-m^{2})v\cdot(p+k)}T(p)\right]\nonumber\\
  =& \int\frac{d^{4}p}{(2\pi)^{4}}{\rm tr}\left[\bar{T}(p)(I_{0}^{T}+I_{1}^{T}v\cdot p+\cdots)T(p)\right],\label{TTselfE}
\end{align}
where we expand the effective Lagrangian in terms of $v\cdot p$. To evaluate the integral we use the proper-time method with a momentum cut off $\Lambda$, and incomplete gamma function
\begin{align}
\frac{1}{k^{2}+m^{2}}  =\int_{1/\Lambda^{2}}^{\infty}ds\ e^{-s(k^{2}+m^{2})},~~~~\Gamma(z,\lambda) =\int_{\lambda}^{\infty}dt\ e^{-t}t^{z-1}.
\end{align}
The coefficients in Eq.~(\ref{TTselfE}) read as
\begin{align}
I_{0}^{T} & =\frac{N_{c}m^{2}}{16\pi^{2}m_{X}}\left[\sqrt{\pi}\Gamma\left(-\frac{1}{2},\frac{m^{2}}{\Lambda^{2}}\right)+\Gamma\left(-1,\frac{m^{2}}{\Lambda^{2}}\right)\right],\nonumber\\
I_{1}^{T} & =\frac{N_{c}m}{8\pi^{2}m_{X}}\left[\Gamma\left(0,\frac{m^{2}}{\Lambda^{2}}\right)+\frac{\sqrt{\pi}}{2}\Gamma\left(-\frac{1}{2},\frac{m^{2}}{\Lambda^{2}}\right)\right].
\end{align}
Finally we arrive at the kinematic term of $T$
\begin{align}
{\cal L}_{T}^{kin} & ={\rm tr}\left[I_{1}^{T}\bar{T}(iv\cdot\partial)T+\left(I_{0}^{T}-\frac{1}{G_{1}}\right)\bar{T}T\right]={\rm tr}\left[\bar{T_{r}}(iv\cdot\partial-\delta m_{T})T_{r}\right].
\end{align}
where we have renormalized the $T$ field as $T=\sqrt{Z_{T}}T_{r}$,
and defined its mass as $\delta m_{T}$. Note that $\delta m_{T}$ is the residual mass of the $T$ baryon, which can be understood as the difference between the baryon mass and the diquark mass $M_T-m_X$. $Z_{T}$ and $\delta m_{T}$ read as
\begin{align}
Z_{T} =(I_{1}^{T})^{-1},~~~~
\delta m_{T} & =Z_{T}\left(\frac{1}{G_{1}}-I_{0}^{T}\right).\label{TMass}
\end{align}
The renormalization of $H_{\mu}$ is similar. The quadratic terms of $H_{\mu}$ come from $-(1/i){\rm Tr}~{\rm log}{\cal D_{\mu\nu}}$ and $(1/G_1) \bar{H}_{ij}^{\mu}H_{ji\mu}$, which leads to the kinematic term
\begin{align}
{\cal L}_{H}^{kin} & ={\rm tr}\left[-I_{1}^{H}\bar{H}_{\mu}(iv\cdot\partial)H^{\mu}-\left(I_{0}^{H}-\frac{1}{G_{1}}\right)\bar{H}_{\mu}H^{\mu}\right]={\rm tr}\left[-\bar{H}_{(r)\mu}(iv\cdot\partial-\delta m_{H})H_{(r)}^{\mu}\right],
\end{align}
where the divergent coefficients are the same as those of $T$: $I_{0,1}^{H}=I_{0,1}^{T}$. After making the redefinition $H^{\mu}=\sqrt{Z_{H}}H_{r}^{\mu}$, one has
\begin{align}
Z_{H} =(I_{1}^{H})^{-1}=Z_{T},~~~~
\delta m_{H} =Z_{H}\left(\frac{1}{G_{1}}-I_{0}^{H}\right)=\delta m_{T}.\label{HMass}
\end{align}

\section{Effective interactions of doubly heavy baryon}
\subsection{EM interaction and Ward-Takahashi identity}

\begin{figure}
\includegraphics[width=0.9\columnwidth]{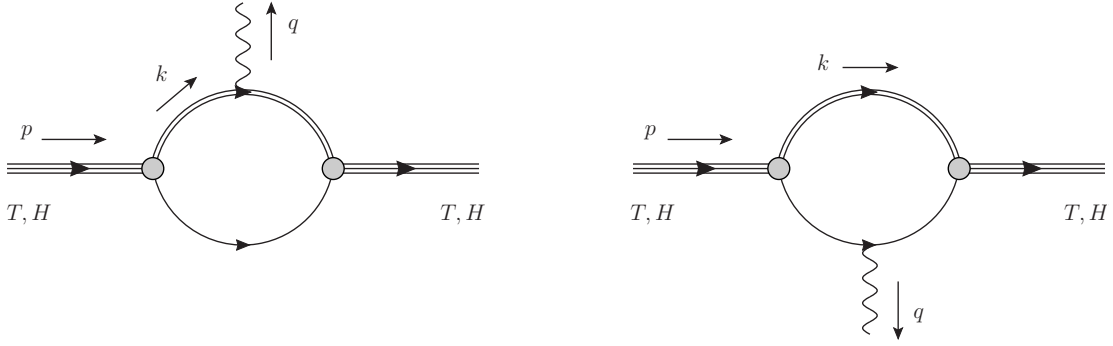} 
\caption{EM interaction of doubly heavy baryon $T$ or $H$, where the photon can be emitted from the heavy diquark (left) or the light quark (right). }
\label{fig:doubBaryonEM} 
\end{figure}

In this section, we firstly consider the effective EM interaction of doubly heavy baryon, and also check the Ward-Takahashi identity at hadron level. Considering the $H-\gamma-H$ vertex, which comes from $-(1/i){\rm Tr}~{\rm log}{\cal D_{\mu\nu}}$. As shown in Fig.~\ref{fig:doubBaryonEM}, there are two corresponding diagrams. In the first one the photon is emitted from the heavy diquark, while in the second one the photon is emitted from the light quark. These two EM vertexes come from the conserved EM currents at diquark and quark level as shown in Eq.~(\ref{Lagran01}) respectively. Although at each level it is ensured that the Ward-Takahashi identity is implied from current conservation, at hadron level this is not obvious and one should check it by perturbative calculations. 

For the first diagram in Fig.~\ref{fig:doubBaryonEM}, we have
\begin{align}
 & i\ {\rm Tr}\ {\rm log}\left[-m_{X}g^{\mu\nu}\delta_{jk}\delta_{il}i v\cdot\partial-2m_{X}g^{\mu\nu}(v\cdot\tilde{E})_{ji,lk}-H_{ji}^{\mu,b}\left(i\slashed D_{em}+m\right)_{ab}^{-1}\bar{H}_{lk}^{\nu,a}\right]\nonumber \\
= & -\frac{2i}{m_{X}}{\rm Tr}\left[(iv\cdot\partial)^{-1}H_{ji}^{\mu,b}(i\slashed\partial+m)_{ab}^{-1}\bar{H}_{mn}^{\nu,a}(iv\cdot\partial)^{-1}(v\cdot\bar{A})_{nm,lk}\right]+\cdots\nonumber \\
= &\  \frac{i}{m_{X}}\int\frac{d^{4}p}{(2\pi)^{4}}\frac{d^{4}q}{(2\pi)^{4}}{\rm tr}\left[\bar{H}_{\mu}(p-q)I_{1}^{EM}(p,q)\{v\cdot\bar{A}(q),H^{\mu}(p)\}\right]+\cdots,\label{EMLagrdiag1}
\end{align}
where the ellipses denote the irrelevant terms with the first diagram. The loop integration is 
\begin{equation}
I_{1}^{EM}(p,q)=\int\frac{d^{4}k}{(2\pi)^{4}}\frac{1}{\slashed k-\slashed p+m}\frac{1}{v\cdot k}\frac{1}{v\cdot(k-q)}.
\end{equation}
Since in the HDiET, the large momentum factor ${\rm exp}[-i\ m_X v\cdot x]$ has been removed, the above integration can be expanded in terms of the external small momentums $p,q$. However, such expansion will contribute to extra power corrections of ${\cal O}(1/m_X)$, which will not be discussed in this work. Therefore, we only need to calculate $I_{1}^{em}(0,0)$ where external momentums vanish. On the other hand, since all the UV divergence are carried by $I_{1}^{em}(0,0)$, it is enough for checking the Ward-Takahashi identity. Thus we have
\begin{equation}
I_{1}^{EM}(0,0)=N_{c}\frac{i}{8\pi^{2}}m\left[\Gamma\left(0,\frac{m^{2}}{\Lambda^{2}}\right)+\frac{\sqrt{\pi}}{2}\Gamma\left(-\frac{1}{2},\frac{m^{2}}{\Lambda^{2}}\right)\right].\label{EMdiag1}
\end{equation}
The contribution from the second diagram in Fig.~\ref{fig:doubBaryonEM} is
\begin{align}
 & \frac{i}{m_{X}}{\rm Tr}\left[(iv\cdot\partial)^{-1}H_{ji}^{\mu,b}\left((i\slashed\partial+m)^{-1}{\slashed E}_{q}(i\slashed\partial+m)^{-1}\right)_{ab}\bar{H}_{mn}^{\nu,a}\right]\nonumber \\
= & -\frac{i}{m_{X}}\int\frac{d^{4}p}{(2\pi)^{4}}\frac{d^{4}q}{(2\pi)^{4}}{\rm tr}\left[\bar{H}_{\mu}(p-q)I_{2,\rho}^{em}(p,q)\bar{A}_{q}^{\rho}(q)H^{\mu}(p)\right],\label{EMLagrdiag2}
\end{align}
where the loop integral reads
\begin{equation}
I_{2,\rho}^{EM}(p,q)=\int\frac{d^{4}k}{(2\pi)^{4}}\frac{1}{v\cdot k}\frac{1}{\slashed k-\slashed p+m}\gamma_{\rho}\frac{1}{\slashed k-\slashed p+\slashed q +m}.
\end{equation}
Similarly, we only contract the UV divergent part, which is
\begin{equation}
I_{2,\rho}^{EM}(0,0)=-N_{c}\frac{i}{8\pi^{2}}m\left[\Gamma\left(0,\frac{m^{2}}{\Lambda^{2}}\right)+\frac{\sqrt{\pi}}{2}\Gamma\left(\frac{1}{2},\frac{m^{2}}{\Lambda^{2}}\right)+\frac{3\sqrt{\pi}}{4}\Gamma\left(-\frac{1}{2},\frac{m^{2}}{\Lambda^{2}}\right)\right]\gamma_{\rho}.\label{EMdiag2}
\end{equation}

Although regularizing the UV divergence by momentum cut off is useful for obtaining effective action, actually it may break the gauge invariance so that is not suitable to verify Ward-Takahashi identity. Therefore Eq.~(\ref{EMdiag1}) and Eq.~(\ref{EMdiag2}), as well as the self energy correction $I_1^H$ should be replaced by the versions in dimensional regularization
\begin{align}
I_{1}^{EM}(0,0)_{dim} & =N_{c}\frac{2\pi^{d/2}i}{(2\pi)^{d}}m^{1-2\epsilon}\left[\Gamma\left(\epsilon\right)+\frac{\sqrt{\pi}}{2}\Gamma\left(\epsilon-\frac{1}{2}\right)\right]=i\ m_X I_{1,dim}^H~,\nonumber \\
I_{2,\rho}^{EM}(0,0)_{dim} & =-N_{c}\frac{2\pi^{d/2}i}{(2\pi)^{d}}m^{1-2\epsilon}\left[\Gamma\left(\epsilon\right)+\frac{\sqrt{\pi}}{2}\Gamma\left(\epsilon-\frac{1}{2}\right)+\frac{\sqrt{\pi}}{4}\epsilon\Gamma\left(\epsilon-\frac{1}{2}\right)\right]\gamma_{\rho}~,
\end{align}
where $d=4-2\epsilon$. It is obvious that in the limit $\epsilon \to 0$, the above two terms are equivalent except the $-\gamma_{\rho}$ factor.  Also note that in the heavy baryon limit, we can simplify the gamma matrix between ${\bar H}^{\mu}$ and $H_{\mu}$ as baryon velocity $v$. These desirable results enable one to combine Eq.~(\ref{EMLagrdiag1}) and Eq.~(\ref{EMLagrdiag2}) to obtain an effective EM interaction for doubly heavy baryon at the leading order expansion  of the external momentums, and the effective EM interaction reads
\begin{align}
{\cal L}_{H\gamma H}^{eff}&=-I_{1}^{H}{\rm tr}\left[\bar{H}_{\mu}\{v\cdot\bar{A},H^{\mu}\}\right]-I_{1}^{H}{\rm tr}\left[\bar{H}_{\mu}\bar{\slashed A}_{q}H^{\mu}\right]\nonumber \\
& =-I_{1}^{H}(v\cdot A^{em})\ {\rm tr}\left[\bar{H}_{\mu}^{a}\{{\cal Q}_{h},H_{a}^{\mu}\}+\bar{H}_{\mu}^{a}{\cal Q}_{l}^{ab}H_{b}^{\mu}\right]\nonumber \\
 & =-(v\cdot A^{em})\sum_{ija}(Q_{i}+Q_{j}+q_{a})\bar{H}_{(r)\mu}^{ij,a}H_{(r)ij,a}^{\mu},\label{HHLeadOrEM}
\end{align}
where, $Q_{i}, Q_{j}, q_{a}$ are the electric charge carried by the three quarks in the baryon. However, at this order one can pretend not to know these internal details, while only treat the value $Q_{i}+Q_{j}+q_{a}$ as the total electric charge carried by the point like baryon. In the last step $I_{1}^{H}$ cancels with the field renormalization factor $Z_H$ after we redefine $H^{\mu}=Z_H H_{(r)}^{\mu}$, which means no further renormalization constants need to be introduced and thus the Ward-Takahashi identity is verified. 

In addition, it should be noted that at higher orders of $p, q$, there is no simple combination for Eq.~(\ref{EMLagrdiag1}) and Eq.~(\ref{EMLagrdiag2}). Thus at higher order one cannot obtain an effective Lagrangian only depending on the total electric charge of the baryon such as Eq.~(\ref{HHLeadOrEM}). In fact, this is understandable since when the external momentums become large, they will probe deeply into the internal of the baryon so that approximating the baryon as a point like particle is no longer valid.

Finally, it can be found that there is no EM interaction of $T-\gamma - H$ or $H-\gamma - T$. For the case of $T-\gamma - T$ interaction, we have a similar result
\begin{align}
{\cal L}_{T\gamma T}^{eff}&=(v\cdot A^{em})\sum_{ija}(Q_{i}+Q_{j}+q_{a})\bar{T}_{(r)}^{ij,a}T_{(r)ij,a}.\label{TTLeadOrEM}
\end{align}

\subsection{Flavor changing process and Isgur-Wise function}

Next, we study the flavor changing process of doubly heavy baryons, which is induced by the coupling with axial vector field $A_{\mu}$. We firstly consider the effective interaction of the form $H-A-H$, which comes from the trace
term
\begin{align}
 & i\ {\rm Tr}{\rm \ log}\left[-m_{X}g^{\mu\nu}\delta_{jk}\delta_{il}i\ v\cdot\partial-H_{ji}^{\mu,b}\left(i\slashed\partial+m\right)_{ab}^{-1}\bar{H}_{lk}^{\nu,a}+2m_{X}\epsilon_{\beta}^{\ \nu\alpha\mu}iv_{\alpha}\tilde{A}_{ji,lk}^{\beta}\right]\nonumber \\
= & \frac{2i}{m_{X}}{\rm Tr}\left[(iv\cdot\partial)^{-1}H_{ji}^{\mu,b}\left(i\slashed\partial+m\right)_{ab}^{-1}\bar{H}_{lk}^{\nu,a}(iv\cdot\partial)^{-1}\epsilon_{\beta}^{\ \nu\alpha\mu}iv_{\alpha}\tilde{A}_{ji,lk}^{\beta}\right]+\cdots\nonumber \\
= & \frac{2}{m_{X}}v^{\rho}\epsilon_{\sigma\nu\rho\mu}Z_H{\rm tr}\left[\bar{H}_{(r)}^{\mu}I^{EW}(0,0)\langle A^{\sigma},H_{(r)}^{\nu}\rangle_{+}\right]+\cdots. \label{HHLeadOrEW}
\end{align}
where $\langle A,B\rangle_{\pm}=(\lambda_1/{2})A\ B\pm(\lambda_2/{2})B\ A^{T}$, and only the relevant terms of the form $H-A-H$ in the trace are retained. The corresponding Feynman diagram is the same as the left one in Fig.~\ref{fig:doubBaryonEM}. It can be found that $I^{EW}(p.q)=I_1^{EM}(p,q)$. Thus the cancelation between $I^{EW}(0,0)$ and $Z_H$ implies that the axial current of $H$ baryon also satisfies the Ward-Takahashi identity. This result is understandable because the diquark axial current as shown in Eq.~(\ref{Lagran01}) is conserved. Similarly, for $T-A-H$ and $H-A-T$ interactions, we have
\begin{equation}
\frac{2i}{m_{X}}\sqrt{Z_{T}}\sqrt{Z_{H}}\left\{{\rm tr}\left[\bar{T}_{(r)}I^{EW}(0,0)\langle A_{\mu},H_{(r)}^{\mu}\rangle_{+}\right]+{\rm tr}\left[\bar{H}_{(r)}^{\mu}I^{EW}(0,0)\langle A_{\mu},T_{(r)}\rangle_{+}\right]\right\}.
\end{equation}

However, for flavor changing processes, where velocities of the initial and final baryons are different, namely $v$ and $v^{\prime}$, $I^{EW}(p.q)$ will also depend on $w=v\cdot v^{\prime}$. To include both of the two velocities, one should insert the diquark flavor changing current in the diagram, and such current has been given in Ref.~\cite{Shi:2020qde}. In fact, this is equal to replace $v_{\rho}$ in Eq.~(\ref{HHLeadOrEW}) with $(v_{\rho}+v_{\rho}^{\prime})/2$ and changing one of the $(iv\cdot\partial)^{-1}$ s to be $(iv^{\prime}\cdot\partial)^{-1}$. As a result, $I^{EW}(p.q)$ becomes $w$ dependent
\begin{align}
I^{EW}(p,q,w) & =\int\frac{d^{4}k}{(2\pi)^{4}}\frac{1}{\slashed k-\slashed p+m}\frac{1}{v\cdot k}\frac{1}{v^{\prime}\cdot(k-q)},\nonumber \\
I^{EW}(0,0,w) & =N_{c}\frac{i}{8\pi^{2}}m\left[r(w)\Gamma\left(0,\frac{m^{2}}{\Lambda^{2}}\right)+\frac{\sqrt{\pi}}{w+1}\Gamma\left(-\frac{1}{2},\frac{m^{2}}{\Lambda^{2}}\right)\right],
\end{align}
with
\begin{equation}
r(w)=\frac{{\rm log}\left[\sqrt{w^{2}-1}+w\right]}{\sqrt{w^{2}-1}}.
\end{equation}
At last, we obtain the effective interaction for coupling with axial vector as
\begin{equation}
{\cal L}_{EW}^{eff}=\xi(w)\ {\rm tr}\Big[i(v^{\rho}+v^{\prime\rho})\epsilon_{\mu\nu\sigma\rho}\bar{H}_{v^{\prime}(r)}^{\mu}\langle A^{\sigma}, H_{v(r)}^{\nu}\rangle_{+}-\bar{T}_{v^{\prime}(r)}\langle A_{\mu},H_{v(r)}^{\mu}\rangle_{+}-\bar{H}_{v^{\prime}(r)}^{\mu}\langle A_{\mu},T_{v(r)}\rangle_{+}\Big],
\end{equation}
where $\xi(w)$ is the Isgur-Wise function, which satisfies $\xi(1)=1$ and reads as
\begin{equation}
\xi(w)=Z_{H}I^{EW}(0,0,w)=\frac{I^{EW}(0,0,w)}{I^{EW}(0,0)}.\label{xiwFunc}
\end{equation}
In addition, the spinor-vector $H_{\mu}$ combines a multiplet of spin-1/2 Dirac field ${\cal B}_{v}$ and a spin-3/2 Rarita-Schwinger field ${\cal B}_{v}^{\mu}$(do not confused with that in Eq~(\ref{BaryoneffAction})). The decomposition reads
\begin{equation}
H_{v}^{\mu}=\frac{1}{\sqrt{3}}\gamma_{5}(\gamma^{\mu}-v^{\mu}){\cal B}_{v}+{\cal B}_{v}^{\mu}.
\end{equation}
As an example, for the transition of the spin-1/2 baryon ${\cal B}_{v}$,  namely the $\Xi_{QQ^{\prime}}$, the corresponding effective interaction is
\begin{align}
{\cal L}_{HAH}^{EW} =\frac{1}{6}\xi(w)\ 2(1+w){\rm tr}\left[\lambda_{1}\bar{{\cal B}}_{v^{\prime}(r)}\slashed A\gamma_{5}{\cal B}_{v(r)}+\lambda_{2}\bar{{\cal B}}_{v^{\prime}(r)}\gamma_{\sigma}\gamma_{5}{\cal B}_{v(r)}A^{T\sigma}\right],
\end{align}
which is consistent with the transition matrix element reduction for ${\cal B}_{v}$ given in Ref.~\cite{Shi:2020qde}:
\begin{align}
\langle{\cal B}^{1/2(1)}_{cQ}|J^{A}_{\mu}(0)|{\cal B}^{1/2(1)}_{bQ}\rangle= \eta(w)\bar{u}\big[2(1+w)\gamma_{\mu}\big]\gamma_{5}u~,\label{FF12A12}
\end{align}
where $J^{A}_{\mu}=\bar c \gamma_{\mu}\gamma_{5} b$, and $\eta(w)=(1/6)\xi(w)$. Note that as proved in Ref.~\cite{Shi:2020qde}, the Isgur-Wise function $\xi(w)$ is proportional to a universal soft function defined by a matrix element only of gluon and light quark fields, which is blind to whatever the heavy sector is and reads
\begin{align}
\xi({w})\propto\langle0|T\Big\{W\Big[\begin{array}{c}
0\\
v^{\prime}
\end{array}\Big]_{i^{\prime}}^{i}W^{-1}\Big[\begin{array}{c}
0\\
v^{\prime}
\end{array}\Big]_{j}^{i^{\prime}}W\Big[\begin{array}{c}
0\\
v
\end{array}\Big]_{k^{\prime}}^{j}W^{-1}\Big[\begin{array}{c}
0\\
v
\end{array}\Big]_{l}^{k^{\prime}}q_{i}^{a}(0)\bar{q}_{a}^{l}(0)\Big\}|0\rangle~,
\label{softfunc}
\end{align}
where $W$ is the Wilson line along the direction of the heavy hadron velocity 
\begin{align}
W\Big[\begin{array}{c}
x\\
v
\end{array}\Big]=P\Big\{{\rm exp}\Big[ig\int_{-\infty}^{v\cdot x}ds\ v\cdot A(s)\Big]\Big\}.
\end{align}
Since the normalization of $\xi(w)$ is fixed at $w=1$,  one can conclude that the Isgur-Wise function obtained here must equal to that for the singly heavy meson transitions in the heavy quark limit.  We will compare our result with that of $B \to D$ transition in the section for numerical studies. 

\subsection{Strong coupling of doubly heavy baryon and pion}

Finally, we study the strong coupling of doubly heavy baryon and $\pi$ mesons. To include the pions, we need to introduce the standard NJL model for $SU(2)_{L}\times SU(2)_{R}$ chiral symmetry by the replacement for Eq.~(\ref{Lagran2}): ${\cal L}_{NJL}\to{\cal L}_{NJL}+{\cal L}_{NJL}^{ch}$, where
\begin{equation}
\mathcal{L}_{\mathrm{NJL}}^{ch}=\frac{G}{2}\left[(\bar{q}q)^{2}+\left(\bar{q}i\gamma_{5}\vec{\tau}q\right)^{2}\right],
\end{equation}
with $G=\kappa/12$, and $\vec{\tau}$ are the isospin Pauli matrices \cite{AbuRaddad:2002pw}. This chiral NJL term can also be linearized by introducing auxiliary meson fields $\sigma$ and $\vec{\pi}$, which leads to
\begin{align}
&{\rm exp}\left[i\int d^{4}x\ \left(\bar{q}(i\slashed D_{em}-m)q+{\cal L}_{NJL}^{ch}\right)\right]\nonumber\\
 =& \int{\cal D}\sigma{\cal D}\vec{\pi}\ {\rm exp}\left[\bar{q}\left(i\slashed D_{em}-m-\sigma-i\gamma_{5}\vec{\tau}\cdot\vec{\pi}\right)q-\frac{1}{2G}\left(\sigma^{2}+\vec{\pi}^{2}\right)\right].
\end{align}
Further, one can transform the meson fields to a non-linear parameterization form
\begin{equation}
\Sigma\equiv\sigma+i\gamma_{5}\vec{\tau}\cdot\vec{\pi}=\left(m_{q}+\sigma^{\text{\ensuremath{\prime}}}\right)\exp\left(-\frac{i}{F_{\pi}}\gamma_{5}\vec{\tau}\cdot\vec{\Phi}\right)~~\text{with}~~\vec{\tau}\cdot\vec{\Phi}=\begin{pmatrix}\pi^{0} & \sqrt{2}\pi^{+}\\
\sqrt{2}\pi^{-} & \pi^{0}
\end{pmatrix},
\end{equation}
where $m_q=\langle \sigma \rangle_{0}$ is the vacuum expectation value of $\sigma$. Then, by repeating the derivation from Eq.~(\ref{genetateFunc}) to Eq.~(\ref{BaryoneffAction}), one will arrives at a modified generating functional including the mesons
\begin{align}
Z[0] =& \int{\cal D}T{\cal D}H_{\mu}{\cal D}\sigma^{\prime}{\cal D}\vec{\Phi}\ {\rm Det}\left[i\slashed D_{em}-m-\Sigma\right]{\rm exp}\left[-{\rm Tr}\ {\rm log}{\cal A}^{\prime}-{\rm Tr}\ {\rm log}\left({\cal D_{\mu\nu}^{\prime}}-{\cal C_{\mu}^{\prime}}{\cal A}^{\prime -1}{\cal B_{\nu}^{\prime}}\right)\right]\nonumber \\
 & \times{\rm exp}\left[i\int d^{4}x\left(-\frac{1}{G_{1}}\bar{T}_{ij}T_{ji}+\frac{1}{G_{1}}\bar{H}_{ij}^{\mu}H_{ji\mu}-\frac{1}{2G}\left(\sigma^{\prime}+m_{q}\right)^{2}+\delta\mathcal{L}_{\mathrm{sb}}\right)\right].
 \end{align}
The primed operators ${\cal A^{\prime}, B^{\prime}, C^{\prime}, D^{\prime}}$ are only different from those defined in Eq.~(\ref{fourOpers}) by a replacement $(m+i\slashed D_{em}^{\prime})^{-1}\to (m+i\slashed D_{em}^{\prime}-\Sigma)^{-1}$. $\delta{\cal L}_{sb}$ is the symmetry-breaking mass term \cite{AbuRaddad:2002pw}.

 \begin{figure}
\includegraphics[width=0.8\columnwidth]{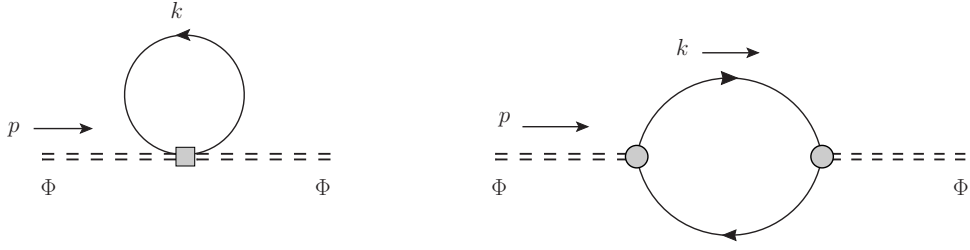} 
\caption{Two diagrams of self energy correction for the meson field, with the meson denoted by double dashed line. Only the second one can contribute to the renormalization factor of the pion field.}
\label{fig:PionSelfE} 
\end{figure}
Note that now the determinant ${\rm Det}\left[i\slashed D_{em}-m-\Sigma\right]$
contains meson fields so that it will contribute to the self energy correction of meson, and the relevant terms are
\begin{equation}
{\cal L}_{\Phi\Phi}\subset i\ {\rm Tr}\left[(i\slashed\partial-m)^{-1}\Sigma\right]+\frac{i}{2}{\rm Tr}\left[(i\slashed\partial-m)^{-1}\Sigma(i\slashed\partial-m)^{-1}\Sigma\right].
\end{equation}
The first trace contributes to the self energy correction corresponding to the first diagram in Fig.~\ref{fig:PionSelfE} , while the second trace corresponds to the second diagram. However, to calculate the strong coupling constant of doubly heavy baryon and the pion, we only care about  the field renormalization factor of the pion field, which only comes from the second diagram and reads
\begin{align}
{\cal L}_{\Phi\Phi} & =-\frac{im_{q}^{2}}{F_{\pi}^{2}}{\rm Tr}\left[\gamma_{5}\Phi^{k}(i\slashed\partial-m)^{-1}\gamma_{5}\Phi^{k}(i\slashed\partial-m)^{-1}\right]\nonumber \\
 & =-\frac{im_{q}^{2}}{F_{\pi}^{2}}\int\frac{d^{4}p}{(2\pi)^{4}}\Phi^{k}(p)\int\frac{d^{4}k}{(2\pi)^{4}}{\rm tr}\left[\gamma_{5}\frac{1}{\slashed k-m}\gamma_{5}\frac{1}{\slashed k-\slashed p-m}\right]\Phi^{k}(-p)\nonumber \\
 & =\frac{1}{2}I_{1}^{\Phi}\partial_{\mu}\Phi^{k}\partial^{\mu}\Phi^{k}-\frac{1}{2}I_{0}^{\Phi}\Phi^{k}\Phi^{k},
\end{align}
where the renormalization factor of $\Phi$ is
\begin{equation}
\Phi=\sqrt{Z_{\Phi}}\Phi_{(r)},~~~~Z_{\Phi}^{-1}=I_{1}^{\Phi}=\frac{N_{c}}{4\pi^{2}}\frac{m_{q}^{2}}{F_{\pi}^{2}}\Gamma\left(0,\frac{m^{2}}{\Lambda^{2}}\right).
\end{equation}
For the strong coupling constant of a spin-1/2 doubly heavy baryon ${\cal B}$ and pion, the coupling Lagrangian comes from the following trace term 
\begin{align}
{\cal L}_{H\Sigma H} & =\frac{i}{m_{X}}{\rm Tr}\left[(iv\cdot\partial)^{-1}H_{ji}^{\mu,b}\left((i\slashed\partial+m)^{-1}\Sigma(i\slashed\partial+m)^{-1}\right)_{ab}\bar{H}_{lk}^{\nu,a}\right]\nonumber \\
 & =-\frac{i}{m_{X}}Z_H\sqrt{Z_{\Phi}}\int\frac{d^{4}p}{(2\pi)^{4}}\frac{d^{4}q}{(2\pi)^{4}}{\rm tr}\left[\bar{H}_{(r)\mu}^{i}(p-q)I^{sc}(p.q)\Sigma_{(r)ij}(q)H_{(r)j}^{\mu}(p)\right].
\end{align}
\begin{figure}
\includegraphics[width=0.45\columnwidth]{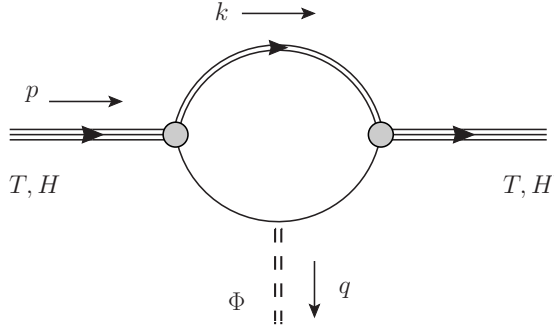} 
\caption{Strong coupling of the doubly heavy baryon $T$ or $H$ and meson $\Phi$.}
\label{fig:doubBaryonSC} 
\end{figure}
The corresponding Feynman diagram is shown in Fig.~\ref{fig:doubBaryonSC}, and the loop integration is
\begin{align}
I^{sc}(p,q) & =\int\frac{d^{4}k}{(2\pi)^{4}}\frac{1}{v\cdot k}\frac{1}{\slashed k-\slashed p+m}\frac{1}{\slashed k-\slashed p+\slashed q+m},\nonumber \\
I^{sc}(0,0) & =\frac{iN_{c}m}{16\pi^{2}}\left[\sqrt{\pi}\Gamma\left(-\frac{1}{2},\frac{m^{2}}{\Lambda^{2}}\right)-2\sqrt{\pi}\Gamma\left(\frac{1}{2},\frac{m^{2}}{\Lambda^{2}}\right)-2\Gamma\left(0,\frac{m^{2}}{\Lambda^{2}}\right)\right],
\end{align}
Here we simply set the external momentum to be zero since this will not affect the strong coupling constant. And finally, we can arrive at the effective strong interaction Lagrangian at leading power as well as the strong coupling constant
\begin{align}
{\cal L}_{{\cal B}\pi{\cal B}} & =-I^{sc}(0,0)Z_{H}\sqrt{Z_{\Phi}}\frac{m_{q}}{F_{\pi}m_{X}}\left(\bar{{\cal B}}_{bb(r)}^{u}\gamma_{5}\pi_{(r)}^{0}{\cal B}_{bb(r)}^{u}+\sqrt{2}\bar{{\cal B}}_{bb(r)}^{u}\gamma_{5}\pi_{(r)}^{+}{\cal B}_{bb(r)}^{d}+\cdots\right),\nonumber\\
g_{{\cal B}\pi^{0}{\cal B}}&=\frac{N_{c}}{16\pi^{2}}\frac{m_{q}m}{F_{\pi}m_{X}}Z_{H}\sqrt{Z_{\Phi}}\left[\sqrt{\pi}\Gamma\left(-\frac{1}{2},\frac{m^{2}}{\Lambda^{2}}\right)-2\sqrt{\pi}\Gamma\left(\frac{1}{2},\frac{m^{2}}{\Lambda^{2}}\right)-2\Gamma\left(0,\frac{m^{2}}{\Lambda^{2}}\right)\right], \label{strongCop}
\end{align}
and $g_{{\cal B}\pi^{\pm}{\cal B}}=\sqrt{2} \ g_{{\cal B}\pi^{0}{\cal B}}$.

\section{Numerical Results}

For the numerical studies in this work, the choice of all the parameters is shown as follows. The quark  masses are set as $m_b=4.18$~GeV, $m_c=1.27$~GeV and $m=(m_u+m_d)/2=3.45$~MeV being the current mass; The  masses of doubly heavy baryons are chosen as $M_{bb}=10.143$~GeV, $M_{bc}=6.943$~GeV and $M_{cc}=3.621$~GeV \cite{Karliner:2014gca,Shah:2016vmd,Shah:2017liu,Kiselev:2001fw}; $\kappa=11.54$~GeV$^{-2}$ \cite{Ebert:1985kz}; $F_{\pi}= 0.093$~GeV, $m_q=0.39$~GeV and the cutoff is set as $\Lambda=0.63$~GeV, which is fixed to yield the constituent quark mass through the NJL gap equation in the meson sector \cite{AbuRaddad:2002pw,Ebert:1985kz,Hatsuda:1994pi,Ebert:1994mf}.

We firstly study the diquark mass and its binding energy, as well as the diquark coupling constant with gluon. We will focus on the case where the two heavy quarks in the baryon form a spin-1 state, which is widely believed in the literatures. According to Eq.~(\ref{HMass}), if such doubly heavy baryon is formed by the hadronization of a heavy axial-vector diquark and a light quark, it will get a residual mass $\delta m_ H$. Explicitly, it can be written as
\begin{align}
\delta m_ H=M_{baryon}-m_X=M_{baryon}-(m_{Q}+m_{Q^{\prime}}+\Delta_X),
\end{align}
where $M_{baryon}$ is the mass of the doubly heavy baryon, $m_{Q}$ and $m_{Q^{\prime}}$ are the masses of the two heavy quarks, while $\Delta_X$ is the binding energy of the heavy diquark.
Since the residual mass $\delta m_ H$ depends on the NJL coupling $G_1$, which is also related with the undetermined coupling constant $g=g_d/g_s$ by the relation $G_{1}=(\kappa/2)m_{X}g$, one can obtain a relationship between $g$ and $\Delta_X$ which is described by the following equation
\begin{align}
Z_{H}\left[\frac{2}{\kappa(m_{Q}+m_{Q^{\prime}}+\Delta_X)g}-I_{0}^{H}\right]=M_{baryon}-(m_{Q}+m_{Q^{\prime}}+\Delta_X).\label{gdAndDeltaX}
\end{align}
\begin{figure}
\includegraphics[width=0.45\columnwidth]{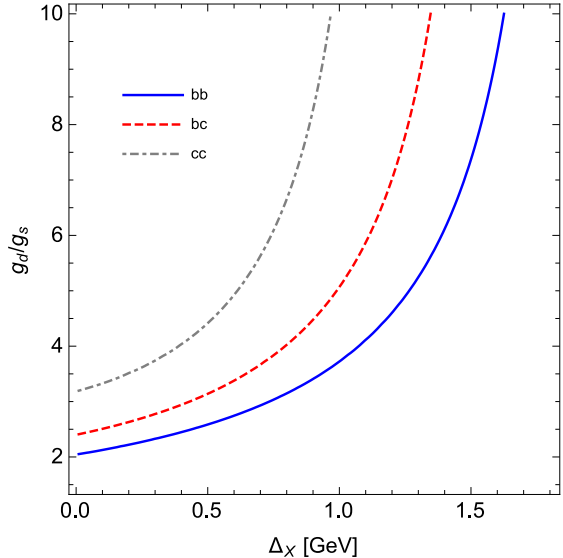} 
\caption{The relationship between $g$ and $\Delta_X$ for ${bb} \text{(blue solid)},\  {bc}\text{(red dashed)}$ and ${cc}\text{(gray dot-dashed)}$ diquarks respectively.}
\label{fig:DiquarkMass} 
\end{figure}
Fig.~\ref{fig:DiquarkMass} shows the curves of this relationship for the ${bb}, {bc}$ and ${cc}$ diquarks respectively. To obtain the value of $g_d/g_s$, one must fix a point on one of the three curves. Here we choose the curve of $bb$ diquark because the heavy diquark limit applied in this work can be safely guaranteed by the heavy bottom mass. The binding energies of various heavy diquarks were obtained by the relativistic quark model in Ref.~\cite{Ebert:2007rn}, where $\Delta_{bb}=1.42$~GeV. Accordingly, using Eq.~(\ref{gdAndDeltaX}) one can get $g_d/g_s=6.33$. Then inserting this coupling value back to Eq.~(\ref{gdAndDeltaX}), we obtain the binding energy of the $bc$ and $cc$ diquarks:
\begin{align}
\Delta_{bc}=1.14~{\rm GeV}~~ \text{and}~~\Delta_{cc}=0.77~{\rm GeV},
\end{align}
which are consistent with those give in Ref.~\cite{Ebert:2007rn}:
\begin{align}
\Delta_{bc}=1.08~{\rm GeV}~~ \text{and}~~\Delta_{cc}=0.69~{\rm GeV}.
\end{align}

\begin{figure}
\includegraphics[width=0.6\columnwidth]{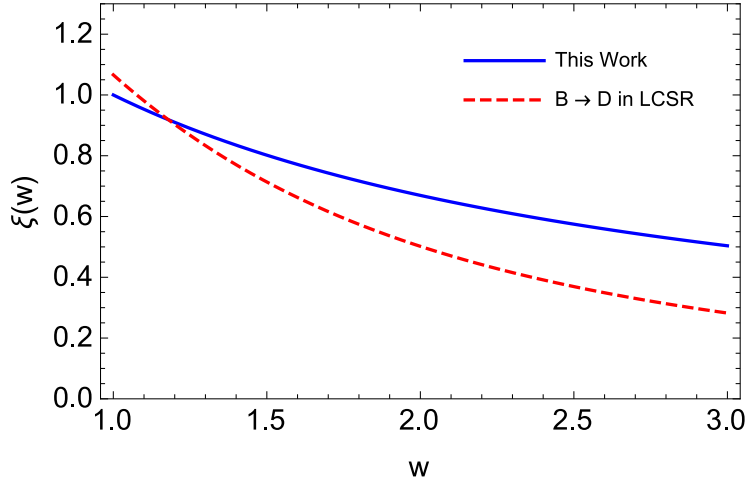} 
\caption{Isgur-Wise function for doubly heavy baryon weak transition given in this work (blue solid line), as well as that for $B \to D$ transition from LCSR calculation (red dashed line).}
\label{fig:xiwFunc} 
\end{figure}
The Isgur-Wise function for doubly heavy baryon weak transition is given in Eq.~(\ref{xiwFunc}), and its curve is shown in Fig.~\ref{fig:xiwFunc}, where we also plot the Isgur-Wise function from the LCSR calculation for the $B \to D$ transition \cite{Faller:2008tr}. As argued in Ref.~\cite{Shi:2020qde}, they  are equivalent due to their common origin from the soft function Eq.~(\ref{softfunc}) in the heavy diquark or quark limit.  However, Fig.~\ref{fig:xiwFunc} shows that the two Isgur-Wise functions are consistent with each other near $w=1$, while have certain difference with larger $w$. This deviation is mainly because that we have only considered the leading power of HDiET, and the higher power studies will be performed in the future works.

Finally, from Eq.~(\ref{strongCop}) the calculation of the strong coupling constants of doubly heavy baryon and pions is straightforward, which read
\begin{align}
g_{{\cal B}\pi^{0}{\cal B}}=1.08~~\text{and}~~ g_{{\cal B}\pi^{\pm}{\cal B}}=1.52,
\end{align}
with $\cal B$ being $\Xi_{bb}$, $\Xi_{bc}$ or $\Xi_{cc}$ baryons.
 
\section{Conclusions}
\label{sec:conclusions}

In summary, we have performed a path-integral hadronization for doubly heavy baryons. The two heavy quarks in the baryon are approximated as a scalar or axial-vector diquark described by a heavy diquark effective theory. The gluon dynamics are represented by a NJL-Model interaction for the heavy diquarks and light quarks, which leads to an effective action of the baryon fields at the leading power of HDiET after the quark and diquark fields are integrated out. We achieved a relationship between the diquark strong coupling $g_d$ and its binding energy $\Delta_X$. We used the binding energy of the $bb$ diquark to obtain that $g_d/g_s=6.33$, with which we then predicted the binding energy of the $bc$ and $cc$ diquarks, and the results are consistent with those from the relativistic quark model. The effective action for doubly heavy baryon derived in this work includes the electromagnetic and electroweak interactions, as well as the interaction with light mesons. For the electromagnetic interaction we proved the Ward-Takahashi identity at the baryon level, while we also pointed out that this verification may become invalid at higher power of HDiET.  For the electroweak interaction we obtained the Isgur-Wise function for weak transitions, and also compared it with that of $B \to D$ transitions. Finally we calculates the strong coupling constant of the doubly heavy baryon and pion, and the result is $g_{{\cal B}\pi^{0}{\cal B}}=1.08$ and  $g_{{\cal B}\pi^{\pm}{\cal B}}=1.52$.

\section*{Acknowledgements}
The author is very grateful to Prof. Zhen-Xing Zhao and Dr. Chien-Yeah Seng for useful discussions. This work is
supported in part by 
the DFG and the NSFC through funds provided to the Sino-German CRC 110 ``Symmetries and the Emergence of
Structure in QCD''.

\end{document}